\documentclass[12pt]{article}
\usepackage{graphicx}

\newcommand\pubnumber{}
\newcommand\pubdate{\today}

\def\fermi{Fermi National Accelerator Laboratory}

\def\Title#1{\begin{center} {\Large #1 } \end{center}}
\def\Author#1{\begin{center}{ \sc #1} \end{center}}
\def\Address#1{\begin{center}{ \it #1} \end{center}}

\newcommand\pubblock{\rightline{\begin{tabular}{l} \pubnumber\\
         \pubdate  \end{tabular}}}
\newenvironment{Abstract}{\begin{quotation}  }{\end{quotation}}
\newenvironment{Presented}{\begin{quotation} \begin{center} 
             PRESENTED AT\end{center}\bigskip 
      \begin{center}\begin{large}}{\end{large}\end{center} \end{quotation}}

\def\beq{\begin{equation}}
\def\eeq#1{\label{#1}\end{equation}}
\def\eeqn{\end{equation}}

\def\beqa{\begin{eqnarray}}
\def\eeqa#1{\label{#1}\end{eqnarray}}
\def\eeqan{\end{eqnarray}}

\let\bar=\overbar

\def\Dslash{\not{\hbox{\kern-4pt $D$}}}
\def\dslash{\not{\hbox{\kern-2pt $\del$}}}

\def\msb{{\bar{\ssstyle M \kern -1pt S}}}

\begin{document}
\begin{titlepage}
\pubblock

\vfill
\Title{Time-integrated measurements of $\gamma$ at the Tevatron and prospects}
\vfill
\Author{ Paola Squillacioti (for the CDF Collaboration)}
\Address{\fermi}
\vfill
\begin{Abstract}
The measurement of CP-violating asymmetries and branching ratios of $B \to DK$ modes allows a theoretically-clean extraction of the CKM angle $\gamma$. We report recent CDF measurements with Cabibbo suppressed ($\pi\pi$, $KK$) or doubly Cabibbo suppressed ($K^+ \pi^-$) $D$ decays. These measurements are performed for the first time in hadron collisions.
\end{Abstract}
\vfill
\begin{Presented}
6th International Workshop on the CKM Unitarity Triangle\\
University of Warwick, UK, September 6-10, 2010
\end{Presented}
\vfill
\end{titlepage}
\def\thefootnote{\fnsymbol{footnote}}
\setcounter{footnote}{0}

\section{Introduction}
\begin{figure}[!h]
 \begin{center}
\includegraphics[scale=0.35]{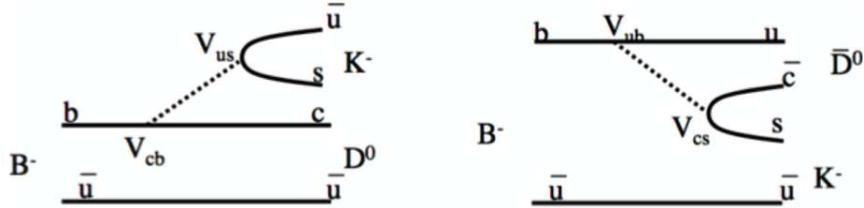}
\caption{Diagrams contributing to $B \to DK$ and related modes. The diagram on the left proceeds via $V_{cb}$ transition, while the diagram on the right proceeds via $V_{ub}$ transition and it is color suppressed.}
\label{diagrams}
\end{center}	
\end{figure}

Using $B\to D K$ decays $\gamma$ could be extracted by exploiting the interference between the tree amplitudes of the $b\rightarrow c\bar{u}s$ and $b\rightarrow u\bar{c}s$ processes. In Fig. \ref{diagrams} the diagrams of these processes are shown, on the left the $B^- \rightarrow {D}^0 K^-$ ($b \rightarrow c \overline{u} s$) and on the right the $B^- \rightarrow \overline{D}^0 K^-$ ($b \rightarrow u \overline{c} s$). $\gamma$ is the relative weak phase between the two diagrams, and in principle can be probed by measuring CP-violating effects in B-decays where the two amplitudes interfere. This can be obtained in several ways, using different choices of $D$ decay channels~\cite{GW,ADS,GGSZ}.

The precision of current experimental data \cite{hfag} is still far from theoretical 
uncertainties and is statistics--limited, so the current 
knowledge of $\gamma$ can be significantly improved by additional experimental measurements.

All mentioned methods for extracting $\gamma$ from $B 
\rightarrow D K$ modes require no tagging or time-dependent 
measurements, and many of them only involve charged particles in 
the final state. They are therefore particularly well--suited to hadron 
collider environment, where the large production can be exploited. 
The use of specialized trigger based on online detection of secondary decay vertexes (SVT trigger \cite{svt}) allow the selection of pure B meson samples.

\section{Atwood-Dunietz-Soni method}\label{ADS_Sec}

We report the first measurement of branching ratios and CP asymmetries of $B\to D_{DCS}K$ modes performed in hadron collisions, based on an integrated luminosity of 5 fb$^{-1}$ collected by CDF. Events where the $D$ meson decays to  the flavor specific mode $K^-\pi^+$ ($D_{CF}$) or the doubly Cabibbo suppressed mode $K^+  \pi^-$ ($D_{DCS}$) are reconstructed.

From these modes, the following observables can be defined \cite{ADS}:
 
\begin{eqnarray}
\displaystyle R_{ADS} & = & \frac{\mathcal{B}(B^-\rightarrow [K^+ \pi^-]_{D}K^-)+\mathcal{B}(B^+\rightarrow [K^-\pi^+]_{D}K^+)}{\mathcal{B}(B^-\rightarrow [K^- \pi^+]_{D}K^-)+\mathcal{B}(B^+\rightarrow [K^+\pi^-]_{D}K^+)}\\
\displaystyle A_{ADS} & = & \frac{\mathcal{B}(B^-\rightarrow [K^+\pi^-]_{D}K^-)-\mathcal{B}(B^+\rightarrow [K^-\pi^+]_{D}K^+)}{\mathcal{B}(B^-\rightarrow [K^+\pi^-]_{D}K^-)+\mathcal{B}(B^+\rightarrow [K^-\pi^+]_{D}K^+)}.
\end{eqnarray}
These quantities are related to the CKM angle $\gamma$ by the equations \cite{ADS}  $R_{ADS}  =  r_D^2 + r_B^2 + 2r_Dr_B \cos{\gamma}\cos{(\delta_B+\delta_D)}$ and 
$A_{ADS}  =  2r_Br_D\sin{\gamma}\sin{(\delta_B+\delta_D)}/R_{ADS}$,
where $r_B$ is the magnitude of the ratio of the amplitudes of the processes $B^-\rightarrow \overline{D}^0 K^-$ and $B^- \rightarrow D^0 K^-$, and $\delta_B$ is their relative strong phase; $r_D$ is the magnitude of the ratio of the amplitudes of the processes $D^0\rightarrow K^-\pi^+$ and $D^0 \rightarrow K^+\pi^-$, and $\delta_D$ is their relative strong phase. We measure $R_{ADS}$ and $A_{ADS}$ also for the $B \to D_{DCS} \pi$ decay mode because also for this mode sizeable asymmetries may be found \cite{hfag}.

\begin{figure}[!h]
 \begin{center}
\includegraphics[scale=0.3]{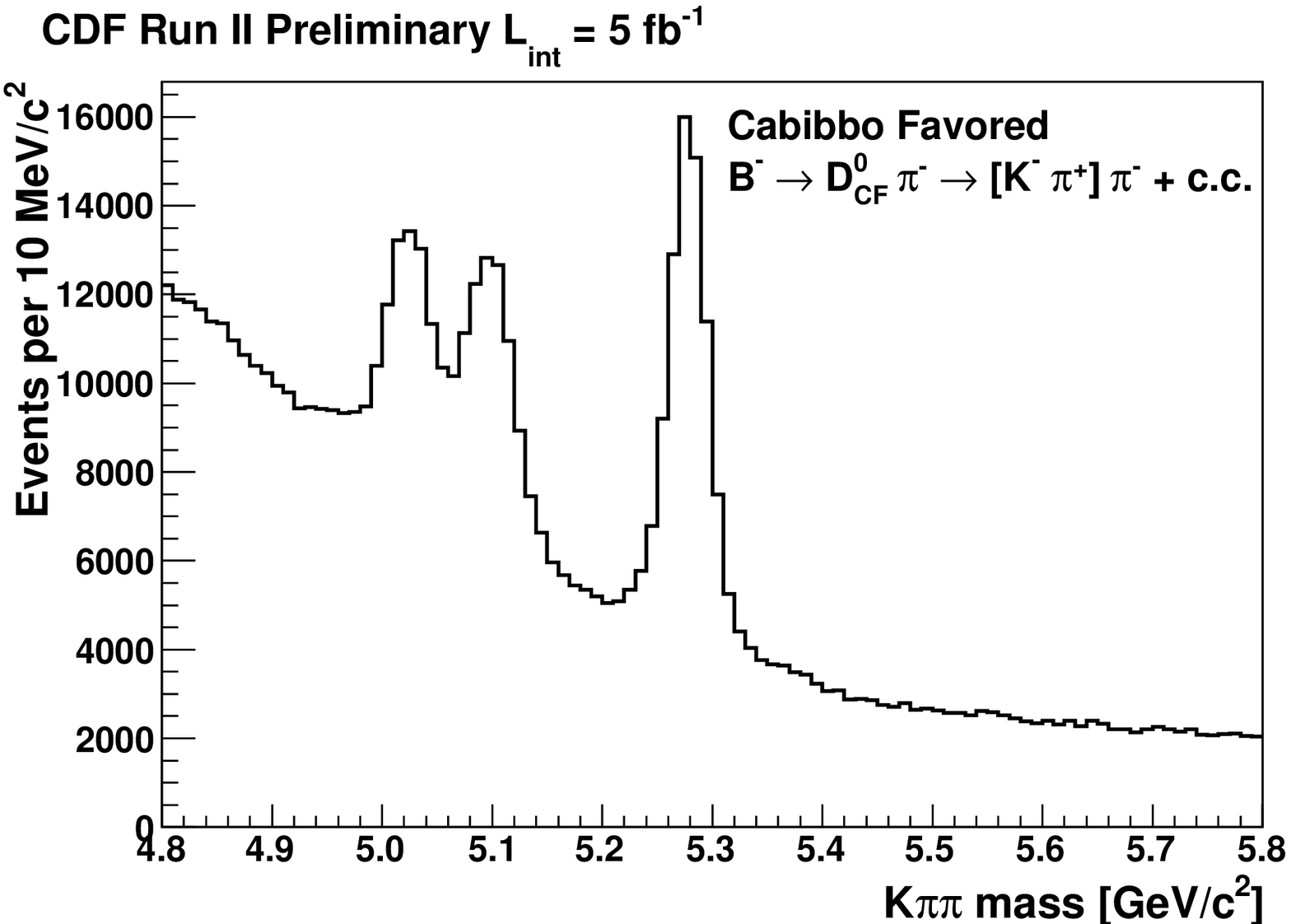}
\includegraphics[scale=0.3]{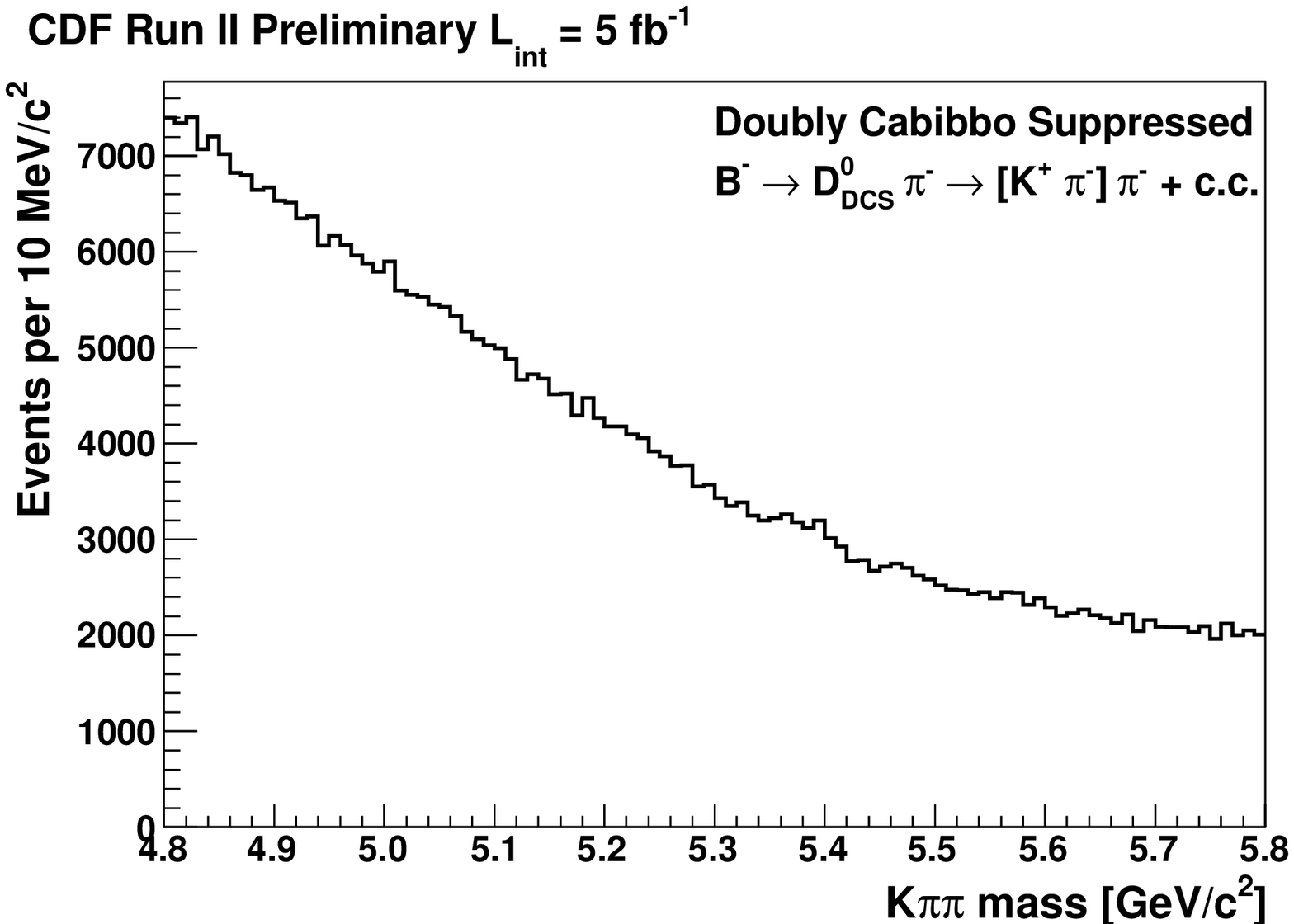}
\caption{Invariant mass distributions of $B \rightarrow D h$
candidates for each reconstructed decay mode. The pion mass is assigned to the track from the $B$ decay.}
\label{ads_bc}
\end{center}
\end{figure}

The invariant mass distributions of CF and DCS modes, with a nominal pion mass assignment
to the track from $B$, are reported in Fig. \ref{ads_bc} where an obvious CF signal is visible, while the DCS signal appears to be buried in the combinatorial background.
Due to the smallness of the DCS branching ratio (0.3\% of the CF rate), the main issue for this analysis is the suppression of the combinatorial background. 
 
\begin{figure}[!h]
 \begin{center}
\includegraphics[scale=0.3]{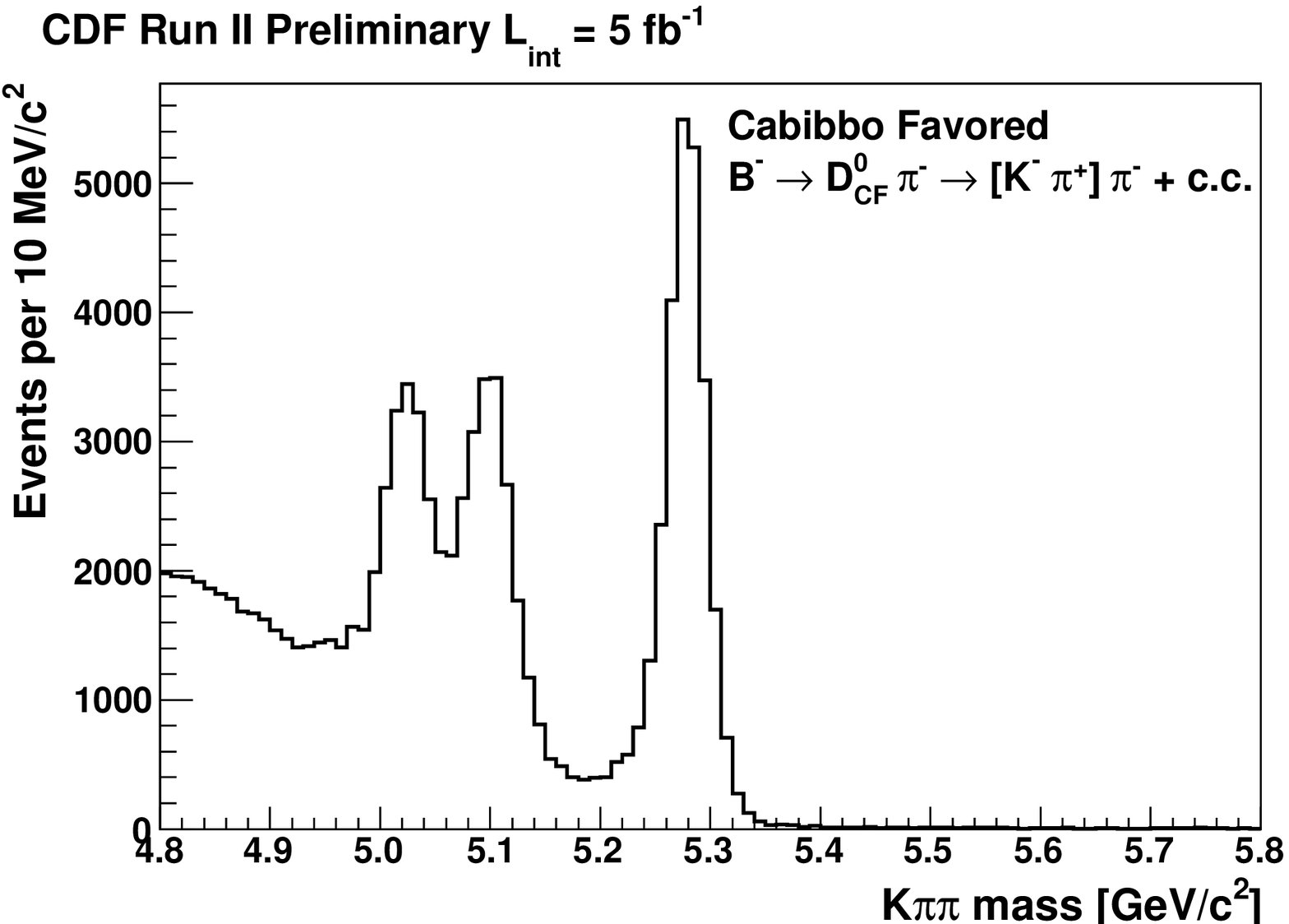}
\includegraphics[scale=0.3]{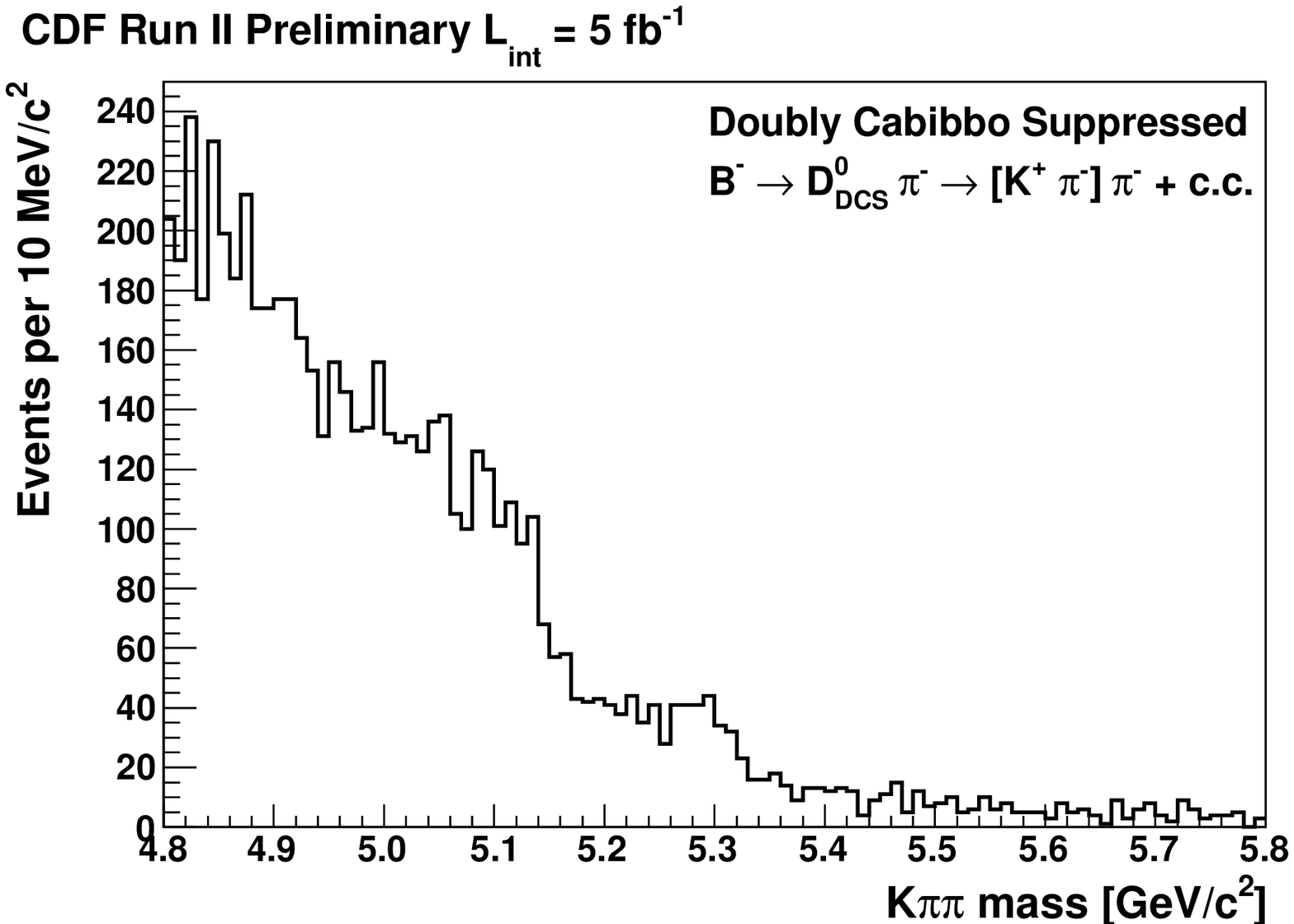}
\caption{Invariant mass distributions of $B \rightarrow D h$
candidates for each reconstructed decay mode after the cuts optimization. The pion mass is assigned to the track from the $B$ decay.}
\label{ads_ac}
\end{center}
\end{figure}

We performed a cuts optimization focused on finding a signal of the $B \to D_{DCS} \pi$ mode.
Since the $B \to D_{CF} \pi$ mode has the same topology of the $DCS$ one, we did the optimization using signal (S) and background (B) from the CF mode. We maximized the figure of merit $S/(1.5 + \sqrt{B})$ \cite{articolo_punzi_sqrt}. The variables used in the optimization and the threshold values for all the requirements are described in \cite{ads_public}.
The resulting invariant mass distributions of CF and DCS modes are reported in Fig. \ref{ads_ac} where the combinatorial background is almost reduced to zero and an indication of the DCS peak is now visible.

The dominant physics backgrounds for the DCS mode are $B^- \to D^0 \pi^-$, with $D^0 \to X$; $B^- \to D^0 K^-$, with $D^0 \to X$; $B^- \to D^{*0} K^-$, with $D^{*0} \to D^0 \gamma/\pi^0$; $B^-Ê\toÊK^-\pi^+\pi^-$ and $B^0 \to D_0^{*-} e^+ \nu_e$ as determined by a study on CDF simulation described in \cite{ads_public}.

An unbinned likelihood fit, exploiting mass and particle identification information provided by the 
specific ionization ($dE/dx$) in the CDF drift chamber, is performed to separate the $B \rightarrow DK$ contributions from the $B \rightarrow D\pi$ signals, from the combinatorial background and from the physics backgrounds.

\begin{figure}[!h]
 \begin{center}
\includegraphics[scale=0.3]{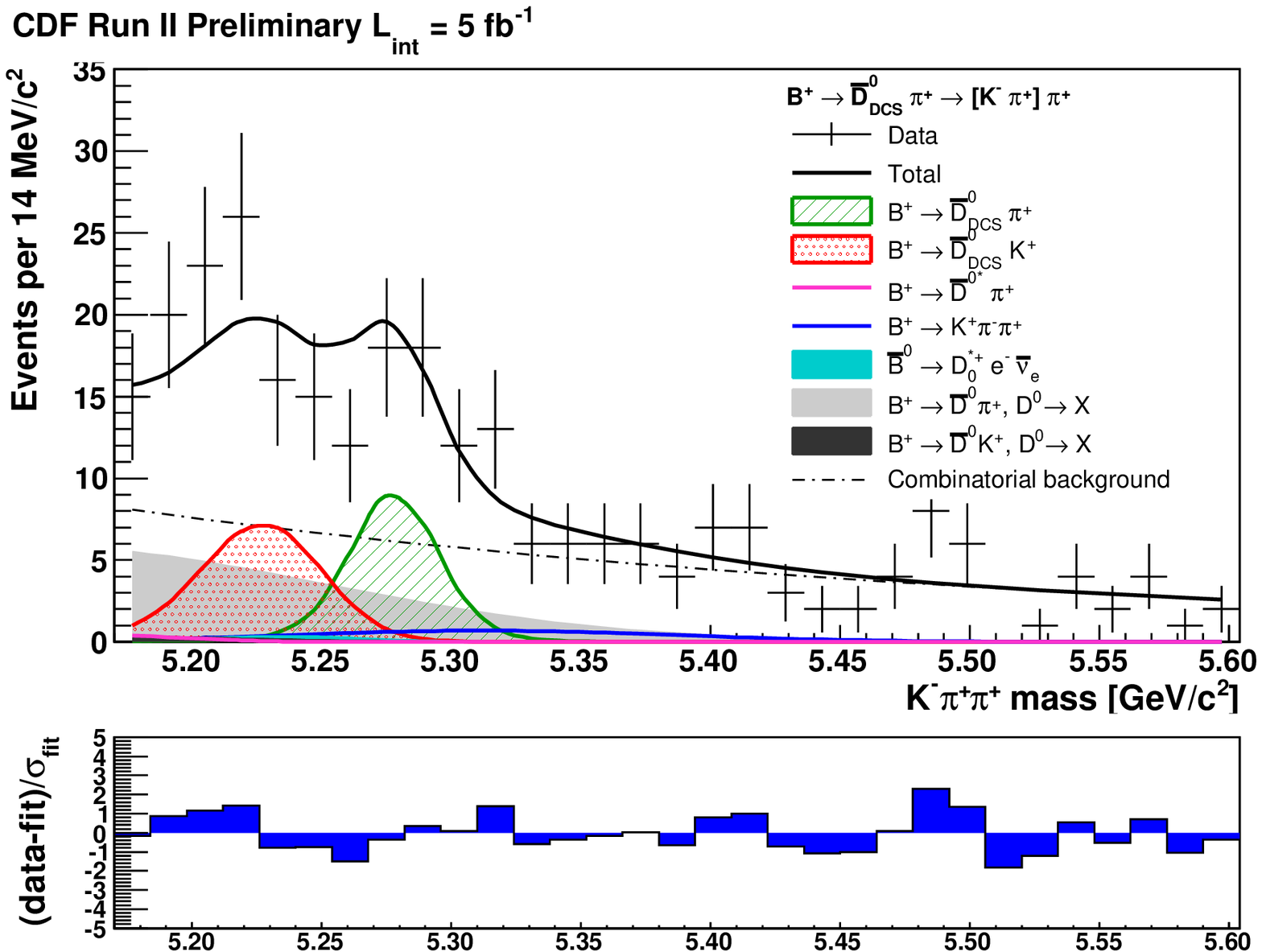}
\includegraphics[scale=0.3]{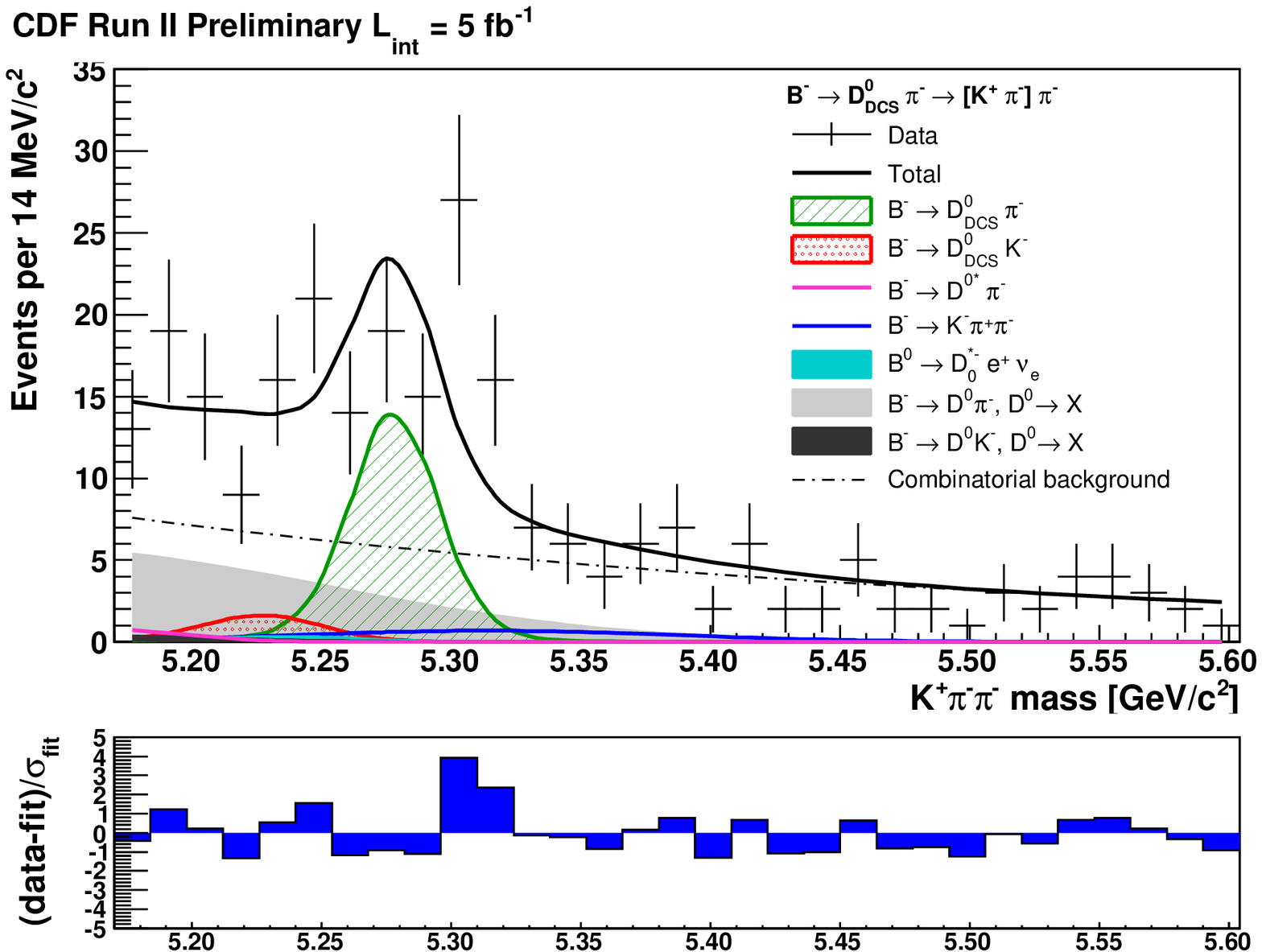}
\caption{Left: Invariant mass distributions of $B^{+} \rightarrow \bar{D}_{DCS} h^{+}$
candidates. Right: Invariant mass distributions of $B^{-} \rightarrow D_{DCS} h^{-}$
candidates. The pion mass is assigned to the track from the $B$ decay. The projections of the likelihood fit are overlaid.}
\label{ads_proj}
\end{center}
\end{figure}

Fig. \ref{ads_proj} shows the DCS invariant mass distributions separated in charge. We obtained $34\pm 14$ $B \to D_{DCS} K$ and $73\pm 16$ $B \to D_{DCS} \pi$ signal events. 

We measured the asymmetries $A_{ADS} (K)= -0.63 \pm 0.40(\rm{stat}) \pm 0.23 (\rm{syst})$ and $A_{ADS} (\pi)= 0.22 \pm 0.18(\rm{stat}) \pm 0.06 (\rm{syst})$ and the ratios of doubly Cabibbo suppressed mode to flavor eigenstate $R_{ADS} (K)= [22.5 \pm 8.4(\rm{stat}) \pm 7.9 (\rm{syst})]\cdot 10^{-3}$ and $R_{ADS} (\pi)= [4.1 \pm 0.8(\rm{stat}) \pm 0.4 (\rm{syst})]\cdot 10^{-3}$.

These quantities are measured for the first time in hadron collisions.
The results are in agreement with existing measurements performed at 
$\Upsilon(4S)$ resonance \cite{hfag,babar_dcs}. 

\section{Gronau-London-Wiler method}\label{GLW_Sec}
\begin{figure}
 \begin{center}
\includegraphics[scale=0.22]{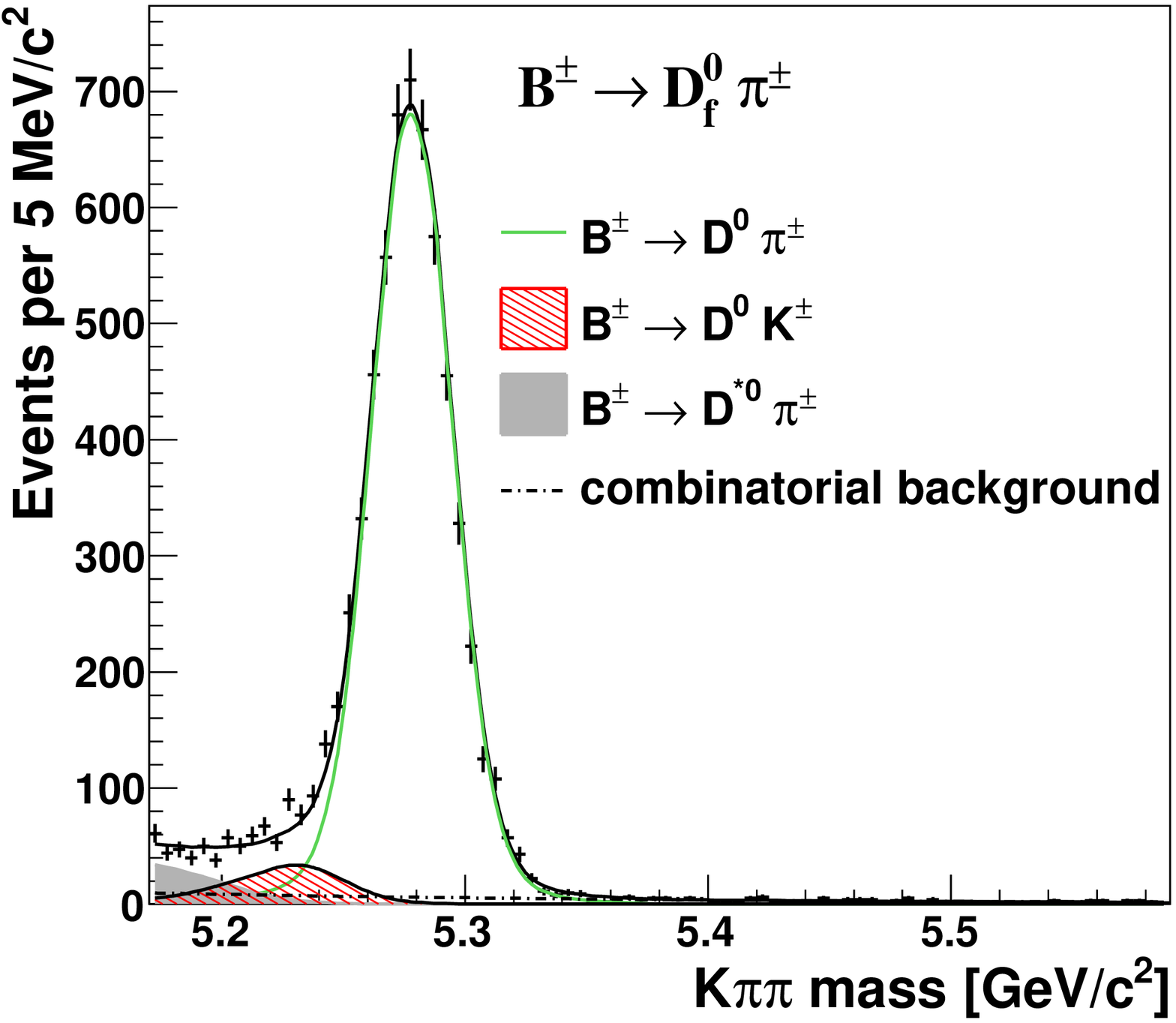}
\includegraphics[scale=0.22]{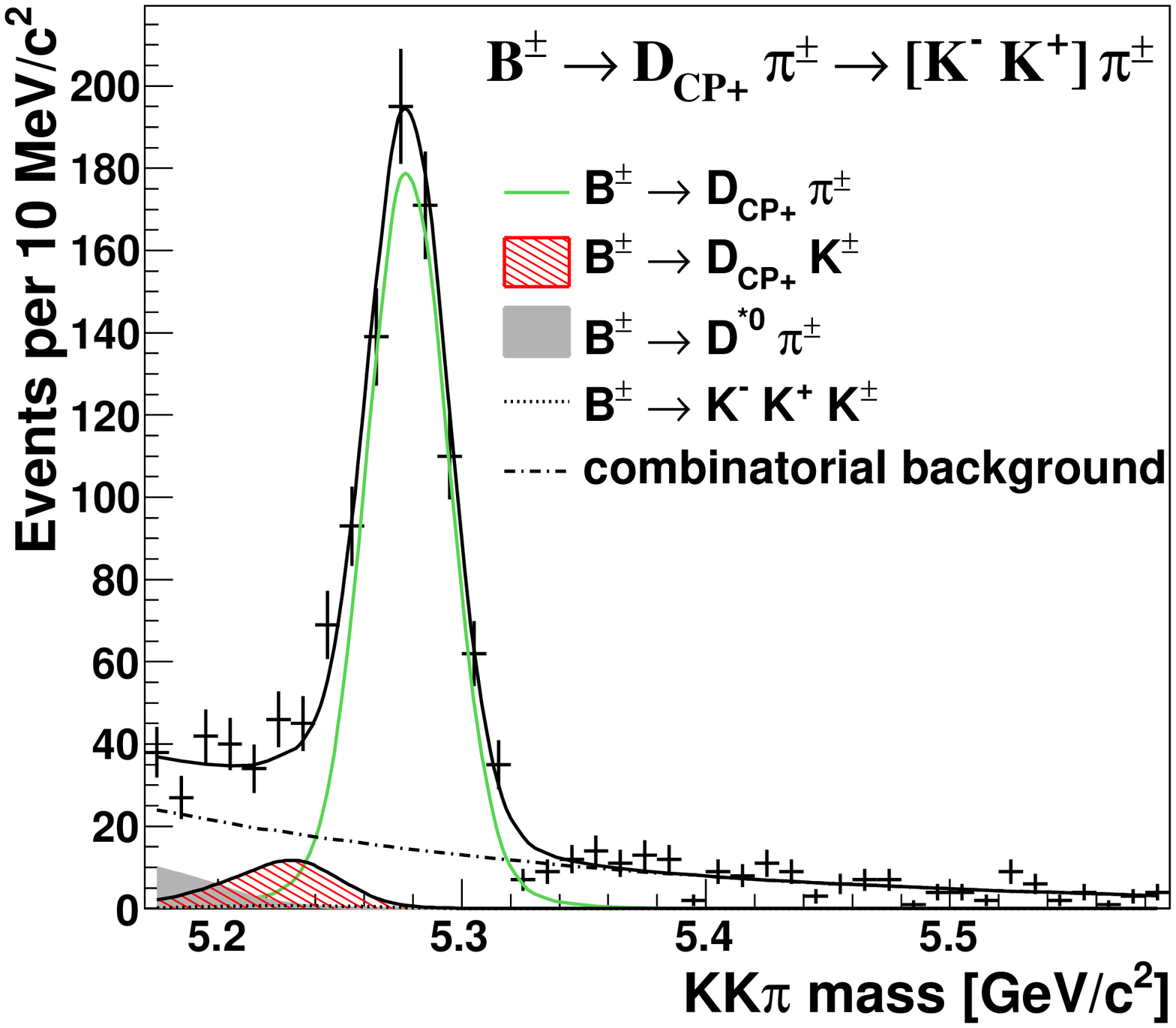}
\includegraphics[scale=0.22]{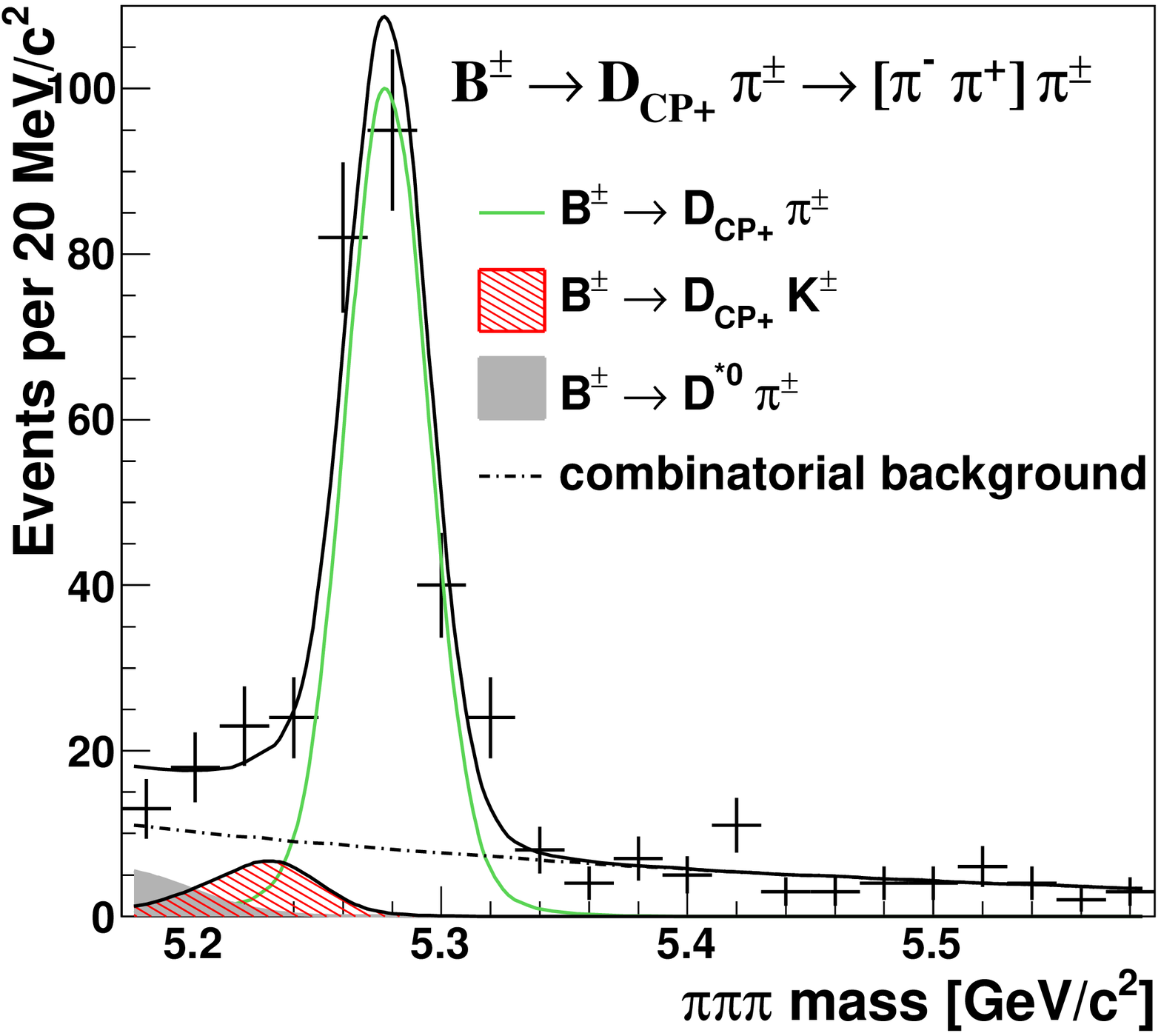}
\caption{Invariant mass distributions of $B \rightarrow D h$
candidates for each reconstructed decay mode. The pion mass is assigned to the track from the $B$ decay. The projections of the likelihood fit are overlaid for each mode.}
\label{dcp}
\end{center}
\end{figure}

We report the first measurement of branching ratios and CP asymmetries of $B\to D_{CP+}K$ modes performed in hadron collisions, based on an integrated luminosity of 1 fb$^{-1}$ collected by CDF \cite{prd_glw}. Events where the $D$ meson decays to the flavor specific mode $K^-\pi^+$, or one of the CP-even modes $K^-K^+$ and $\pi^-\pi^+$ are reconstructed. From these modes, the following observables can be defined~\cite{GW}:
      \begin{equation}\label{acp_eq}
    A_{CP+} = \frac{\mathcal{B}(B^- \rightarrow D_{CP+} K^-)-\mathcal{B}(B^+
    \rightarrow D_{CP+} K^+)}{\mathcal{B}(B^- \rightarrow D_{CP+}
    K^-)+\mathcal{B}(B^+ \rightarrow D_{CP+} K^+)},
    \end{equation}
    \begin{equation}\label{rcp_eq}
    R_{CP+} = 2 \frac{\mathcal{B}(B^- \rightarrow D_{CP+} K^-)+\mathcal{B}(B^+ \rightarrow
    D_{CP+} K^+)}{\mathcal{B}(B^-
 \rightarrow D_{CF} K^-)+\mathcal{B}(B^+ \rightarrow
 \overline{D}_{CF} K^+)}.
  \end{equation}
These quatities are related to the CKM angle $\gamma$ by the equations~\cite{GW} $R_{CP+}=1+r_B^2+2r_B \cos{\delta_B}\cos{\gamma}$ and $A_{CP+}=2r_B \sin{\delta_B \sin{\gamma}/R_{CP+}}$.
For every $B\to Dh$ candidate, a nominal invariant mass is evaluated by assigning the pion mass to the particle $h$ coming from the $B$ decay. The distributions obtained for the three modes of interest (where $D\to K\pi$, $KK$ or $\pi\pi$) are reported in Fig. \ref{dcp}; a clear
$B \rightarrow D \pi$ signal is seen in each.
Events from $B \to D K$ decays are expected to form much smaller 
and wider peaks in these plots, located about 50 MeV/$c^{2}$ below 
the $B \rightarrow D \pi$ peaks, and as such cannot be resolved.
The dominant backgrounds are combinatorial background and mis-reconstructed physics background such as $B\rightarrow D^{*0} \pi$ decay, and in the $D\rightarrow KK$ final state, the non resonant $B\rightarrow K K K$ decay, as determined by a study on CDF simulation.
An unbinned likelihood
fit, exploiting kinematic and particle
identification information provided by the
$dE/dx$, is performed to statistically separate the $B \to D K$ contributions
from the $B\to D \pi$ signals, from the combinatorial
background and from the physics backgrounds.

We obtained around 90 $B \to D_{CP+} K$ and we measured the double ratio of $CP$-even to flavor eigenstate branching fractions $R_{CP+} = 1.30\pm 0.24(\rm{stat})\pm 0.12(\rm{syst})$ and the direct $CP$ asymmetry $A_{CP+} = 0.39\pm 0.17(\rm{stat})\pm 0.04(\rm{syst})$. These results are in agreement with previous measurements from $\Upsilon(4S)$ decays \cite{babar1,belle1}. 

\section{Conclusions}
CDF performed the first measurement of $A_{ADS}$ and $R_{ADS}$ at a hadron collider using a luminosity of 5 fb$^{-1}$. This supplements the recently published first GLW analysis in hadron collisions \cite{prd_glw} within a CDF global program to measure $\gamma$ angle from tree-dominated processes. At the moment we recorded 8 fb$^{-1}$ and we will expect around 10-12 fb$^{-1}$ by the end of 2011. CDF has demonstrated the feasibility of these measurements at a hadron collider and will obtain interesting and competitive results in the near future.
 


\end{document}